%% file: ms.tex
\documentclass[10pt, conference]{IEEEtrans}  %

\IEEEoverridecommandlockouts                              %
\usepackage{blindtext}
\usepackage{amsmath}
\usepackage{amssymb}
\usepackage{commath}
\usepackage{geometry}
\usepackage{listings}
\usepackage{mathtools}
\usepackage{multirow}
\usepackage{siunitx}
\usepackage{algorithm}
\usepackage{algpseudocode}
\usepackage{todonotes}
\usepackage{subcaption}
\usepackage[sorting=none,natbib,backend=bibtex]{biblatex}

\definecolor{darkgreen}{rgb}{0.0, 0.33, 0.0}
\newcommand{\avg}[1]{\overline{#1}}
\newcommand{\predict}[1]{\hat{#1}}

\title{\LARGE \bf
  Model Predictive Congestion Control for TCP Endpoints
}

\author{Taran Lynn$^{1}$ and Dipak Ghosal$^{2}$ and Nathan Hanford$^{3}$%
\thanks{*This research was supported by NSF grant CNS-1528087}%
\thanks{$^{1}$Taran Lynn is an alumnus and assistant researcher at the University
        of California, Davis
    {\tt\small tflynn@ucdavis.edu}}%
\thanks{$^{2}$Dipak Ghosal is with Department of Computer Science at the University
of California, Davis {\tt\small dghosal@ucdavis.edu}}%
\thanks{$^{3}$Nathan Hanford is a postdoctoral fellow at Lawrence Livermore National Laboratory, Livermore, CA {\tt\small nate.hanford@gmail.com}} }%

\begin{document}

\maketitle
\thispagestyle{plain}
\pagestyle{plain}

\begin{abstract}
A common problem in science networks and private wide area networks (WANs) is
that of achieving predictable data transfers of multiple concurrent flows by
maintaining specific pacing rates for each.
We address this problem by developing a control algorithm based on concepts from
model predictive control (MPC) to produce flows with smooth pacing rates and
round trip times (RTTs).
In the proposed approach, we model the bottleneck link as a queue and derive 
a model relating the pacing rate and the RTT.
A MPC based control algorithm based on this model is shown to avoid the extreme
window  (which translates to rate)  reduction that exists in
current control algorithms when facing network congestion.
We have implemented our algorithm as a Linux kernel module.
Through simulation and experimental analysis, we show that our algorithm
achieves the goals of a low standard deviation of RTT and pacing rate, even
when the bottleneck link is fully utilized. In the case of multiple flows, we can assign different rates to each flow and as long as the sum of rates is less than bottleneck rate, they can maintain their assigned pacing rate with low standard deviation. This is achieved even when the flows have different RTTs. 

\end{abstract}

\input{introduction}

\input{implementation}

\input{results}

\input{buffer}

\input{relatedwork}

\input{conclusion}
\printbibliography

\end{document}

%% file: introduction.tex
\section{Introduction}
\label{sec:introduction}

TCP is still the dominant transport-layer protocol on the Internet.
This is in part due to its ability to guarantee reliable delivery even
with a high degree of unpredictability in network quality and speed
\cite{kurose2013computer}.
It additionally benefits from widespread support of offload hardware
for high performance network environments \cite{hanford2018survey}.
This has made it the protocol of choice for private WANs used for
scientific computing, and some private networks like Google's B4
\cite{bwe}, which require large amounts of data to be transferred over
long distances quickly and reliably~\cite{dart2014science}.
These networks are governed by various security policies, such as
Access Control Lists (ACLs), and as a result they are ``quieter''
compared to the chaotic traffic of the open Internet.
There have been multiple methods developed for performing TCP congestion
control.

\subsection{Loss Based Schemes}
\label{sec:intro_loss}

The earliest and most developed congestion control algorithms use packet
loss as their primary feedback.
Examples of such algorithms include TCP Reno \cite{RFC5681}, Cubic \cite{cubic},
BIC \cite{bic}, and H-TCP \cite{htcp}.
These algorithms usually use the congestion window (cwnd) as their
main method of controlling congestion.
That is, they limit the number of unacknowledged packets in the network.
They also tend to use additive increase/multiplicative decrease (AIMD)
techniques for updating the cwnd.
Such techniques increase the cwnd by some number of packets (one for Reno),
and decrease it to a fraction of its original value on a loss (50\% of
the cwnd for Reno).

This approach has a number of advantages.
If a network has appropriately sized buffers, then loss based systems
will use the full bandwidth.
For networks that have very little random packet loss, such as many
wired networks, packet loss tends to be a very robust against network
fluctuations due to factors like packet processing.
It is fairly trivial to prove that AIMD systems are fair when all flows
have the same round trip time (RTT) \cite{chiu1989analysis}.

However, these advantages are offset by two major issues.
First, the scheme perform poorly in networks that experience high
levels of random losses.
Such networks include most all networks with Wi-Fi access points, which
make up an increasing number of access points.
Second, these schemes only react when maximum congestion has occurred,
resulting in lost packets.
At this point the buffers have all been saturated, leading to large
RTTs as packets wait to be processed.
For TCP Reno to achieve full bandwidth throughput this increase, at
minimum, is on average 1.5 times the base RTT.
At worst, large buffers can lead to seconds of latency, crippling
real-time applications like online video games, video conferencing
software, and more.
This effect is known as bufferbloat \cite{bufferbloat}\cite{gettys2011bufferbloat}.

\subsection{Delay Based Schemes}
\label{sec:delay_intro}

Recently developed congestion control algorithms use the RTT of
packets as an indicator for congestion in order to avoid the
pitfalls of loss based systems.
Algorithms that use this scheme include Vegas \cite{vegas}, TIMELY \cite{timely},
BBR \cite{bbr_tcp}, and HPCC \cite{hpcc}.
Some of these schemes (namely TIMELY and BBR) use the pacing rate
(the rate at which data is transmitted) as their main control,
instead of the cwnd (TIMELY) or in addition to it (BBR).
The pacing rate is used because congestion occurs when this rate
exceeds the rate at which the bottleneck link can process packets,
so there exists a close connection between it and congestion.

The main advantages of delay based algorithms is that they alleviate
some of the issues with loss based schemes.
Since the feedback of such algorithms is not based on packet loss,
they tend to be more resilient in networks with high amounts of
random packet loss, such as those with Wi-Fi endpoints.
Using RTT as a measure also has the benefit of lowering network
latency, which can improve the performance of real-time applications.

The main disadvantage of this scheme is that RTT can be a noisy and
coarse signal.
It can be affected by many other factors than congestion, including
router processing, kernel delays on endpoints, etc.
This can make it difficult to differentiate congestion from normal
fluctuations, which can lead algorithms to be more cautious and slow
when adapting to changing network conditions.
TIMELY tries to avoid some of this noise by working directly on the
NIC, but this only handles noise caused by the kernel.
Recently, HPCC has been developed to use network telemetry from
switchs to precisely calculate latency due to congestion.

\subsection{Hybrid Schemes}
\label{sec:intro_hybrid}

Some algorithms incorporate elements of both schemes.
Notable examples are Compound TCP \cite{compound_tcp} and TCP Illinois
\cite{illinois}.
Two main methods of meshing the two schemes together are present in
these algorithms.
Compound TCP computes windows based off of feedback from losses and
RTT separately, and then combines the windows afterwards.
Illinois uses an AIMD approach governed by losses, but has the amount
of addition or multiplicative decrease determined by the RTT.
Each approach trades off advantages and disadvantages of loss based
and delay based schemes.

\subsection{Our Approach}
\label{sec:our_approach}

Our congestion control algorithm uses a delay based scheme
concepts and from model predictive control (MPC) theory
\cite{bordons2007camacho}.
MPC is advantageous over other control strategies (like PID) when
dealing with complex systems \cite{pan2013pie}.
The main advantage is that MPC can enable us to estimate our control
as a continuous function.
This in turn leads to smoother RTTs and transmission rates.
One caveat is that MPC methods require the system's reactions to
correlate directly to the actions taken by the controller, and thereby
also require that stochastic noise to be relatively low.
This makes it suited to the relatively quite dedicated WANs, but not to general
Internet applications (which can be very noisy and unpredictable).
This can lead to improved performance for long-lived, large transfer connections (aka elephant flows).

The main contributions of this paper are the following.
\begin{enumerate}
\item Based on a simple queuing model of the bottleneck link and parameter estimation methods, we have designed a MPC based traffic control algorithm.

\item We have implemented the control algorithm in the Linux kernel both as congestion control at the transport layer as well as a queuing discipline (qDisc) in the network layer packet scheduler.

\item Through simulation and experimental results we demonstrate that our  algorithm achieves the goals of low standard deviation for RTT and pacing rates, while maximally utilizing the bottleneck rate.

\item Our implementation allows per flow rate to be set.
 We show that if this is used to cap flows so that their sum is less than the bottleneck rate, then we can get low standard deviation even among multiple flows that may have different RTTs.
\end{enumerate}

%% file: implementation.tex
\section{Implementation\protect\footnote{Linux source code for the algorithm is available upon request.}}
\label{sec:implementation}

As with all MPC implementations, the goal is to set a variable (called the
\textbf{response}), which we can only indirectly manipulate using a value
(called the \textbf{control}) that we control directly, such that it matches
some desired value (called the \textbf{target value}).
What makes MPC special is that it predicts the response for several time steps
in the future, and tries to optimize the control for each of these time steps to
match the response.
This process is repeated at each time step, and is thus called the
\textbf{receding horizon} (the idea being that the horizon of prediction
recedes into the future as we gain more information).
Below we describe the three core pieces of MPC as they apply to our control algorithm.

\begin{figure}
    \includegraphics[width=0.9\linewidth]{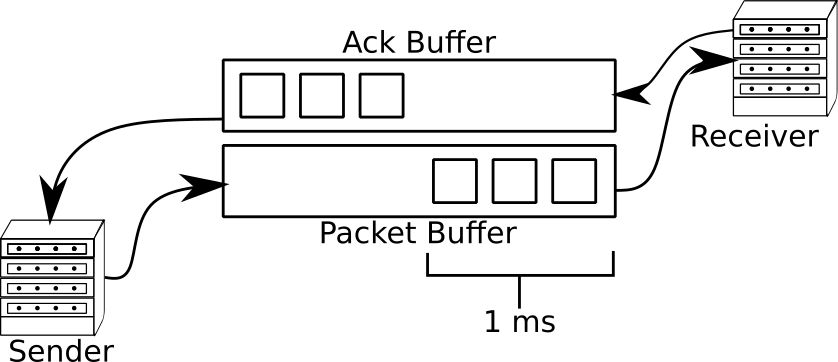}
    \caption{Increasing the pacing rate causes packet buildup in the bottleneck link. This in turn causes an increase in RTT as packets must
    transverse long buffers.}
    \label{fig:buffering}
\end{figure}

\paragraph{Model}
We model the network as a simple queue, and that our round trip time (RTT) increases as the queue becomes larger. Fig. \ref{fig:buffering} shows how increased buffering adds to the total RTT.
From this we can establish several key parameters for the model.
The first is the \textbf{propagation latency} $l_P$, which is the
minimum RTT of the network, and corresponds to an empty queue.
The second is the \textbf{bottleneck rate} $r_B$, which is the pacing
rate that when exceeded leads to an increase in RTT, and corresponds to
processing rate of the queue.
The third is the \textbf{bottleneck latency} $l_B$, which is the maximum
RTT for the network, and corresponds to a full queue.
Taking these three parameters into account the model is
\begin{align}
  \dod{l(t)}{t} &= \frac{r(t) - r_B}{r_B} \label{eq:model_cont} ,
\end{align}
which can be discretized as
\begin{align}
    l(n + 1) &= l(n) + \frac{r(t) - r_B}{r_B} \Delta{t}(n) \label{eq:model_l} .
\end{align}
This is a discrete model indexed by $n$, where $t(n)$ is the time at
index $n$, $\Delta{t}(n) = t(n + 1) - t(n)$, and $r(n)$ and $l(n)$ are
the pacing rate and latency at time $t(n)$.
We must also impose the constraint that $l_P \leq l(n) \leq l_B$ for
all $n$.

\paragraph{Prediction}
To build a prediction we need to estimate the parameters for the
model ($\predict{x}$ denotes the predicted or estimated value of $x$).
For $l_P$ and $l_B$ we simply take the minimum and maximum RTT with exponential
back-off.
The back-off is necessary to account for changing or erroneous minimum and
maximum RTT values.
Exponential back-off allows for quick recovery while maintaining algorithmic
stability.
Mathematically, this corresponds to
\begin{align}
    b_e(x, n) &= x + \frac{\avg{l} - x}{\tau_d} \Delta{t}(n)\\
    \predict{l}_P(n + 1) &= \min\cbr{l(n), b_e(\predict{l}_P(n), n)}\\
    \predict{l}_B(n + 1) &= \max\cbr{l(n), b_e(\predict{l}_B(n), n)} ,
\end{align}
where $b_e$ is the exponential back-off function and $\tau_d$ is the decay time.
If it were in continuous form, we would get
$$b_e(x) = \avg{l} - (\avg{l} - l_P(t_s)) e^{(t_s - t)/\tau_d} ,$$
with $t_s$ being the last time a minimum or maximum RTT was observed.
This shows why the function is an exponential back-off, and that when
$t - t_s = \tau_d$ it will have decayed to 37\% of it's starting value.
To better estimate $l_P$ and $l_B$, we also occasionally decrease and increase
the pacing rate to probe the RTT.

Our first approach to estimating $r_B$ was to use gradient descent to
minimize the error between previous predictions for $l$ and the
measured value of $l$.
However, this approach was unstable, so we instead solved for $r_B$
from equations \ref{eq:model_cont}, giving
\begin{align*}
  \predict{r}_B(t) &= \frac{\int_0^t r(t) \dif{t}}{l(t) - l(0) + t} .
\end{align*}
This formula can be turned into the discrete form
\begin{align}
  \predict{r}_B(n) &= \frac{\sum_{k = 0}^{n - 1} r(k) \Delta{t}(k)}{l(n) - l(0) + t(n) - t(0)} \label{eq:rb_est} .
\end{align}

We then plug these equations into equation \ref{eq:model_l} to get the
predicted latency $\predict{l}(n + 1)$.

\paragraph{Optimization} To optimize the pacing rate we first establish a target
latency $l_t$ (usually set by the user somewhere between $l_P$ and $l_B$).
We then set the rate to minimize the difference between the $\predict{l}$
and $l_t$, the predicted variance in $l$, and the variance of the
pacing rate.
These three things are given the weights $0 < 1 - c_1 - c_2 < 1$,
$0 \leq c_1 \leq 1$, and $0 < c_2 < 1$, respectively.
These weights are set by the user.
The resultant formula is
\begin{align}
    r(n + 1) &= & \frac{c_3 l_P^2 r(n) - \Delta{t}(n) \predict{r_B}(n) \Lambda}{c_3 l_P^2 + \Delta{t}(n)^2 (\alpha c_2 + c_1)}\\ \nonumber
    \Lambda &= & (\alpha c_2 + c_1) \del{l(n) - \Delta{t}(n)} \\
            & &- c_1 l_t - \alpha c_2 \avg{l}(n)
\end{align}
where  $\alpha$ is the weight used for the running average (i.e. we compute
$\avg{l}(n + 1) = (1 - \alpha) \avg{l}(n) + \alpha l(n)$), and is set by
the user (conventionally to $\frac{1}{8}$).
$\Lambda$ is a common value that comes out when deriving the formula,
and represents the RTT part of the optimization.

\subsection{Linux Advanced Routing and Traffic Control and TCP}

\begin{figure}[h]
    \centering
    \includegraphics[width=0.5\textwidth]{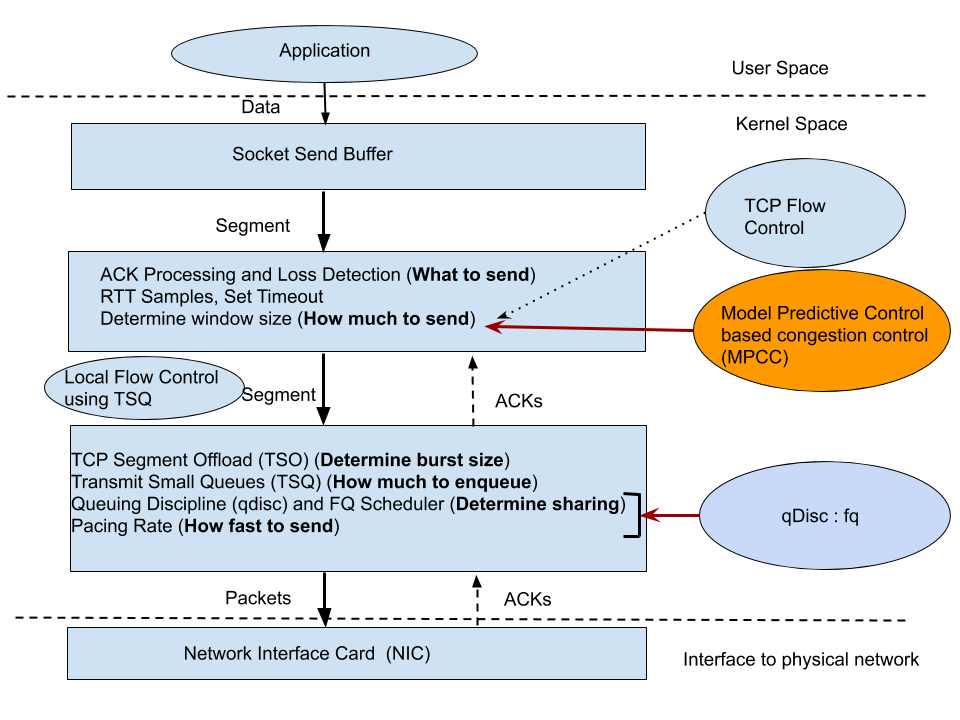}
    \caption{Sender-side functions that determine the rate at which data is injected into the network (adapted from \cite{cheng1making} and \cite{david}). }
    \label{fig:sender_side_functions}
\end{figure}

Figure~\ref{fig:sender_side_functions} shows the sender-side functions that together determine the rate at which the end-system injects data into the network. 
As mentioned in Section~\ref{sec:introduction} we are concerned with large data transfers (elephant flows) and as a result the rate at which data is injected into the network is not application limited.  The application writes into the socket send buffer and it is the function of the  TCP protocol to determine which and how much of the data should be sent. Which data will be sent is determined by the Acknowledgements (ACKs) from the receiver that provide the information as to which data has been received and which has not. 

How much data will be sent, which determines the window size (TCP is a window-based protocol), is determined by two functions namely the TCP flow control and the TCP congestion control algorithms. The end-to-end flow control algorithm allows the receiver to inform the sender how much data it can accept. The receiver does so by indicating available space in its receive socket buffer in a  receiver advertised window size in the ACKs. How much data should be sent based on the network congestion is determined by the congestion control algorithm. There are many TCP congestion control algorithms in use \cite{afanasyev2010host,bbr_tcp,cubic,htcp}. These algorithms determine the congestion window based on the estimate of the  network congestion. The sender uses the minimum of the congestion window and the receiver advertised window sizes to determine how much data can be sent.

The data provided by  TCP  is first processed by  lower level protocols  and then provided to the network interface for transmission. These protocol processing functions determine how much data is eventually transmitted \cite{cheng1making, david}. These functions include TCP Segmentation Offload (TSO), Queuing Discipline (Qdisc) and FQ Packet Scheduling, and Pacing.  The TSO allows TCP to hand off large segments (up to 64 KB) to the lower layer which segments the large chunks into MTU size packets. This improves both the sender-side protocol processing efficiency as well  improved network performance. TSQ enables a local flow control between the network layer queue and the TCP layer. 

The Queuing Discipline (qDisc) and FQ Packet Scheduler determines how the outgoing packets are queued and contains functions to filter and classify packets and methods to handle packets with the same flow descriptor. For fair bandwidth sharing between flows, the scheduler implements a Fair Queue (FQ) mechanism by performing round-robin packet scheduling over a red-black tree which stores the queue state for each flow \cite{cheng1making,esnetoscars}.  The goal of pacing is to spread out the transmission of chunks of data determined by TSO~\cite{HABIBIGHARAKHEILI2015103, wei2006tcp}. This is as opposed to transmitting in a burst as soon as data is made available. In CoDel, pacing is implemented by setting limits on the queue size. Pacing allows rate limits to be enforced. 

As shown in Figure~\ref{fig:sender_side_functions}, the MPC based control algorithm can be implemented either as a  
congestion control or as a qDisc pacing algorithm. These two implementations do
not need to, and are not designed to, operate together. The MPC qDisc can be operated  with
different congestion controls, while the MPC
congestion control may only be operated  with the FQ qDisc.

%% file: results.tex
\section{Results}
\label{sec:results}

In the following subsections we present simulation and experimental results.
We consider two  types of scenarios, namely, uncapped and capped. 
For the uncapped cases we set each flow's maximum rate to be much higher than the bottleneck rate. 
For the capped case we set the maximum rates of each flow to rates that add up to the bottleneck rate.

\subsection{Simulation Results}
\label{sub:simulated_results}

We developed an event based simulation model in which  multiple sources  generated packets at a rate determined by our MPC based control algorithm.
These packets are  placed in a queue that represents the bottleneck link.
Packets are  dequeued at the bottleneck rate and an ACK is delivered to the source after one RTT propagation delay.
The queue has a finite buffer and  losses are recorded when the buffer overflows.
The losses together with the RTT is passed to the MPC algorithm.
Other than the queuing delay at the bottleneck link, a random exponential delay is added to the RTTs  to simulate network noise.

\begin{figure*}[h]
    \begin{subfigure}{.5\textwidth}
    \centering
    \includegraphics[width=0.9\linewidth]{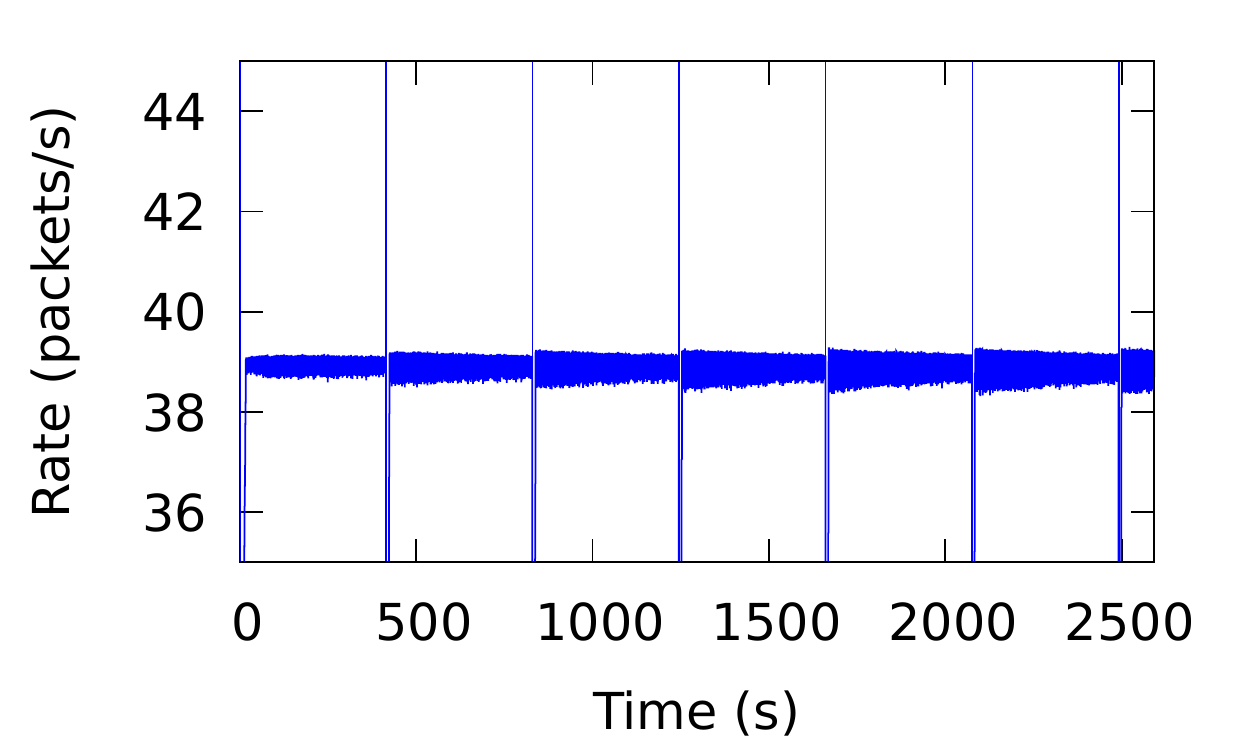}
    \caption{Rate}
    \end{subfigure}
    \begin{subfigure}{.5\textwidth}
    \includegraphics[width=0.9\linewidth]{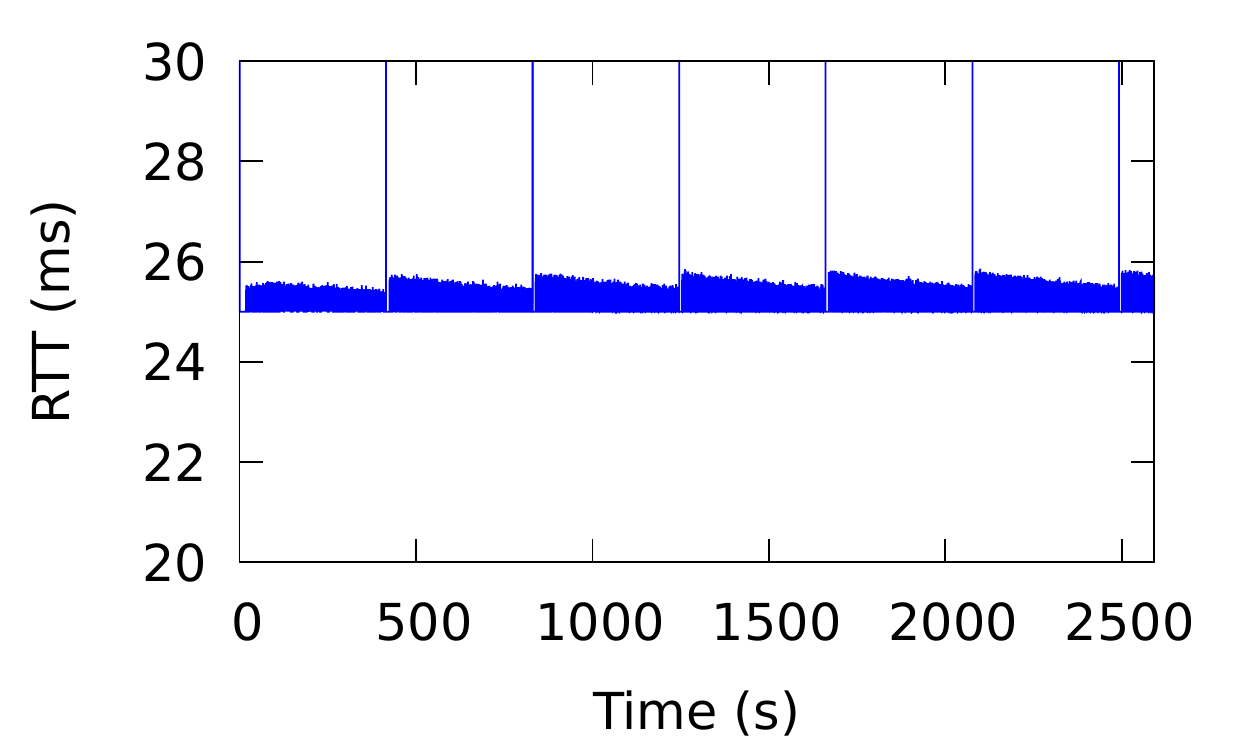}
    \caption{RTT}
    \end{subfigure}
    \caption{Simulation results of (a) achieved rate and (b) RTT for a single flow.
    We can see the lack of a saw tooth pattern that is common in many other TCP
    congestion control algorithms.
    The spikes in the graph correspond to the algorithm probing for $l_P$ and
    $l_B$.}
    \label{fig:sim_single_flow}
\end{figure*}

Figure~\ref{fig:sim_single_flow} shows the simulation results of the achieved rate (throughput) and RTT for a single flow. We observe that the achieved rate and the RTT are very stable (i.e., they have low standard deviation) and they  lack the sawtooth pattern that is common in many other congestion
control algorithms.

When there are multiple flows with the same RTT, the flows in the long run equally share the bottleneck capacity.  This is  shown in Table~\ref{tab:sim_results_multiple_flows_same_rtt} with the columns titled "uncapped." Each of the four flows stay fairly close to the fair rate (the bottleneck rate divided by 4) of 10 packets/s.
They also show a smooth rate and RTT profile similar to the single flow case. When the flows are capped to specific rates,  they achieve that rate with low standard deviation of the rate and the RTT. This is shown in Table~\ref{tab:sim_results_multiple_flows_same_rtt} with the columns titled "capped" where the flows were capped to 3, 7, 10, and 20 packets/s (summing to the bottleneck rate of 40 packets/s)
\begin{table*}[h]
    \centering

    \begin{tabular}{|r|c|c|c|c||c|c|c|c|}
        \cline{2-9}
        \multicolumn{1}{c|}{} & \multicolumn{4}{c||}{Uncapped} & \multicolumn{4}{c|}{Capped} \\
        \hline
        Flow \# &1 & 2 & 3 & 4 & 1 & 2 & 3 & 4\\ \hline
        Mean Rate (packets/s) &9.6 & 9.5 & 9.6 & 10.1 & 3.0 & 7.0 & 10.0 & 18.8\\ \hline
        Rate Std. (packets/s) & 0.1 & 0.1 & 0.1 & 0.3 & 0.0 & 0.1 & 0.1 & 0.2\\ \hline
        Mean RTT (ms) & 25.2 & 25.2 & 25.2 & 25.2 & 25.0 & 25.0 & 25.1 & 25.0\\ \hline
        RTT Std. (ms) & 0.2 & 0.2 & 0.2 & 0.2 & 0.0 & 0.0 & 0.0 & 0.0\\ \hline
    \end{tabular}

    \caption{Simulation results for four flows with the same RTT.
        The bottleneck rate is 40 packets/s.
        For the capped case we limit rates to 3, 7, 10, and 20 packets/s,
        respectively.
        This demonstrates how a network scheduler can selectively limit flows to specific rates.
        Flow 4 has a  higher average pacing rate because when the flows
        initially started it dominated. They converge in the long run.}
    \label{tab:sim_results_multiple_flows_same_rtt}
\end{table*}

Table~\ref{tab:sim_results_diff} shows the results for multiple flows with different RTTs.
If the rates are capped then the flows will maintain their assigned rates with low standard deviation.
This is shown in the capped column.
If the flows are uncapped then one flow will dominate over the others with two
cases.
In the first case the target RTT is set close to $l_P$ and high RTTs flows end up dominating.
In the second case the target RTT is set close to $l_B$ and low RTTs flows end up dominating.
This effect is caused by low RTTs reacting faster than high RTTs, combined with
lower targets reducing flow aggressiveness.
\begin{table*}[h]
    \centering

    \begin{tabular}{|r|c|c|c|c||c|c|c|c|}
        \cline{2-9}
        \multicolumn{1}{c|}{} & \multicolumn{4}{c||}{Uncapped} & \multicolumn{4}{c|}{Capped}\\
        \hline
        Flow \# & 1 & 2 & 3 & 4 & 1 & 2 & 3 & 4\\ \hline
        Mean Rate (packets/s) & 5.8 & 8.0 & 10.5 & 15.3 & 9.0 & 10.0 & 10.0 & 10.0\\ \hline
        Rate Std. (packets/s) & 1.1 & 1.0 & 0.6 & 0.9 & 0.1 & 0.1 & 0.0 & 0.0\\ \hline
        Mean RTT (ms) & 25.2 & 35.2 & 45.2 & 55.2 & 25.0 & 35.0 & 45.0 & 55.0\\ \hline
        RTT Std. (ms) & 0.2 & 0.2 & 0.2 & 0.2 & 0.0 & 0.0 & 0.0 & 0.0\\ \hline
    \end{tabular}

    \caption{Simulation results for four flows with the different
        RTTs.
        The bottleneck rate was 40 packets/s.
        For the capped case we limit rates to 10 packets/s. For the uncapped case the results are unpredictable.}
    \label{tab:sim_results_diff}
\end{table*}

\subsection{Experimental Results}
\label{sub:experimental_results}

For experimental testing, we ran tests over a 10 Gbps link to ESnet's
test servers \cite{esnet}.
The tests were ran on a dedicated WAN where the 10 Gbps link was the  primary
bottleneck.
These tests were made using the pschedular tool, which is part of the perfSonar
suite \cite{tierney2009perfsonar}.
pschedular ensures that we are not in contention for the server before starting
an iPerf3~\cite{mortimer2018iperf3} session to measure network performance.
Target servers were selected to give a broad range of RTTs; and includes servers
in Sacramento, Denver, and New York.
Overall, the results, some of which are discussed below, were similar to those observed from the simulation analysis.

\begin{figure*}[h]
    \begin{subfigure}{.5\textwidth}
    \includegraphics[width=0.9\linewidth]{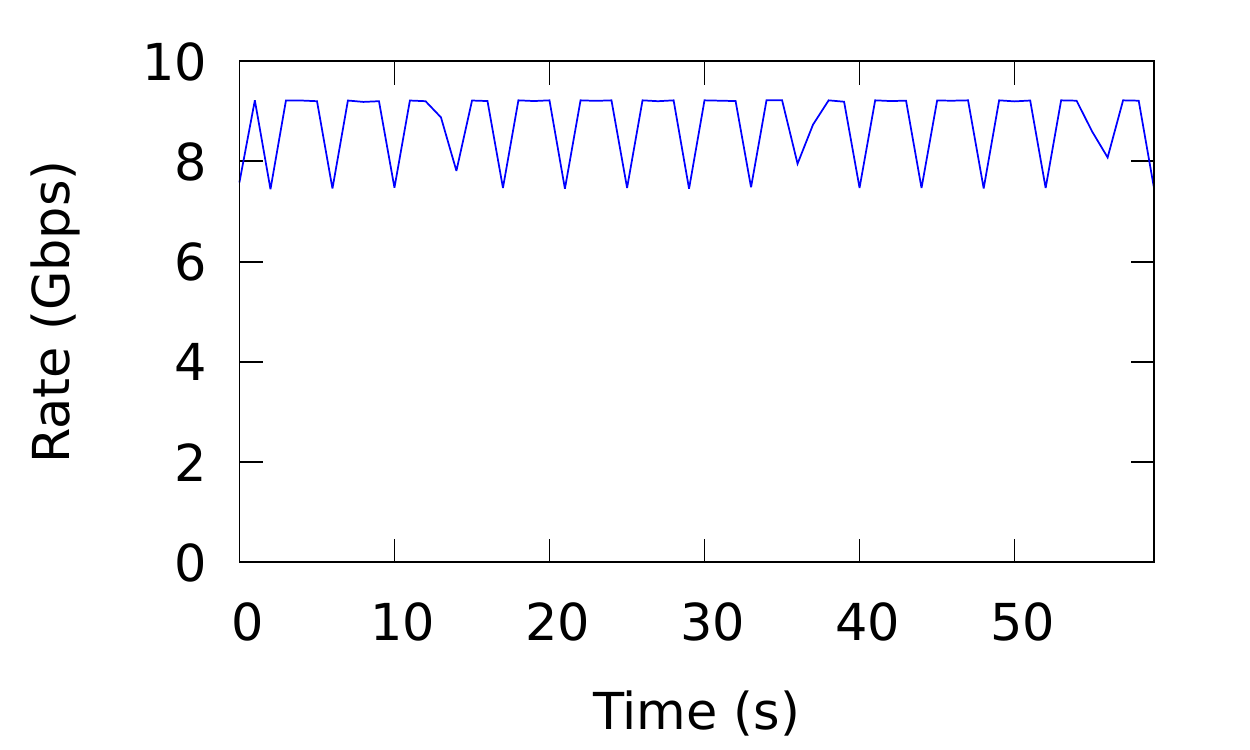}
    \end{subfigure}
    \begin{subfigure}{.5\textwidth}
    \includegraphics[width=0.9\linewidth]{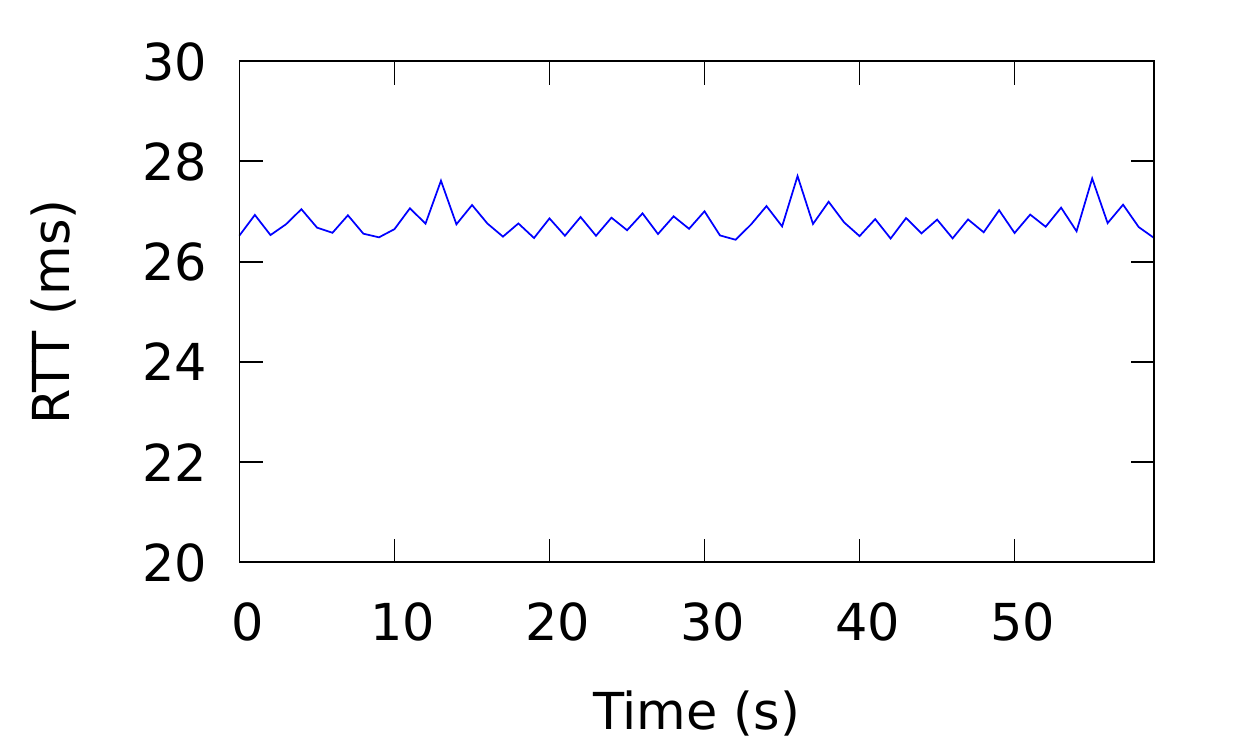}
    \end{subfigure}
    \caption{Experimental results of the achieved rate and the RTT  for a single flow between Davis, CA and Denver, CO. The bottleneck rate was 10~Gbps.
    Results are similar to simulation results. There are slightly more fluctuations in the RTT.  This is due to the relatively higher noise in the physical network.
    The drops in rate correspond to probing for $l_B$.
    Losses are unreported because in all cases there were none.}
    \label{fig:exp_single_flow}
\end{figure*}

When running a single flow between Davis, CA and Denver, CO we achieved results similar
to those in simulation. The results are shown in Fig. \ref{fig:exp_single_flow}.
We achieved a rate of 9.4~Gbps (and not  the full  10 Gbps) because of protocol overhead.  This is also the case for other 
congestion control algorithms such as CUBIC~\cite{cubic}.
However, some small loss of throughput  can be  attributed to down
probing to determine $l_B$.
Nevertheless, the achieved rate and RTT are  stable and do not show a saw tooth
pattern as found in other AIMD-based congestion control algorithms. 

Results using multiple flows to Denver are shown in Table~\ref{tab:exp_results_same}. Since the RTTs are the same,  both the flows equally  share 
bottleneck bandwidth when they are uncapped. This is similar to the simulation results. 
In the case where the flows are capped to a specific rate (3.0 and 6.4 Gbps) the flows achieved those rates with low standard deviation. This is the same as the simulation results. 
When testing different RTTs we used New York, NY as a testing point in addition to
Denver, CO and the results are shown in Table
\ref{tab:exp_results_diff}
We found that Denver would dominate the bandwidth usage, showing that the
algorithm favors lower RTTs.
This is expected, as the algorithm was made more aggressive to counteract the
relative noisiness of the network compared to simulation.
As in simulation, an aggressive algorithm favors lower RTTs.

\begin{table}[h]
    \centering

    \begin{tabular}{|r|c|c||c|c|}
        \cline{2-5}
        \multicolumn{1}{c|}{} & \multicolumn{2}{c||}{Uncapped} & \multicolumn{2}{c|}{Capped}\\
        \hline
        Flow \#          & 1    & 2    & 1    & 2    \\ \hline
        Mean Rate (Gbps) & 4.7  & 4.5  & 2.9  & 6.4  \\ \hline
        Rate Std. (Gbps) & 1.2  & 1.2  & 0.1  & 0.2  \\ \hline
        Mean RTT (ms)    & 27.1 & 27.1 & 27.1 & 26.6 \\ \hline
        RTT Std. (ms)    & 0.4  & 0.5  & 0.2  & 0.1  \\ \hline
        Losses           & 4    & 3    & 2    & 4    \\ \hline
    \end{tabular}

    \caption{Experimental results for two flows between Davis, CA. and Denver CO.
    For the uncapped case we can see that the bottleneck rate of 9.2/9.3 Gbps is (approx.) equally shared. 
    For the capped case we assigned 3.0 Gbps and 6.4 Gbps respectively and the flows were almost able to achieve those rates.
    }
    \label{tab:exp_results_same}
\end{table}

\begin{table}[h]
    \centering

    \begin{tabular}{|r|c|c||c|c|}
        \cline{2-5}
        \multicolumn{1}{c|}{} & \multicolumn{2}{c||}{Uncapped} & \multicolumn{2}{c|}{Capped}\\
        \hline
        Flow \#          & Denver & NY      & Denver & NY\\ \hline
        Mean Rate(Gbps) & 8.6    & 2.3     & 6.9    & 2.5\\ \hline
        Rate Std. (Gbps) & 1.7    & 0.6     & 1.9    & 0.2\\ \hline
        Mean RTT (ms)    & 27.0   & 72.5    & 26.5   & 71.0\\ \hline
        RTT Std. (ms)    & 0.4    & 0.2     & 0.2    & 0.1\\ \hline
        Losses           & 2      & 327,435 & 0      & 2\\ \hline
    \end{tabular}

    \caption{Experimental results for tests from Davis, CA. to Denver, CO. and
    New York, NY.
    Unfortunately, the bottleneck rates to the two destinations were different (10 Gbps to Denver and approx. 4.7 Gbps to New York. These results are different from  simulation, where we had long RTTs dominating.  
    The capped case had the Denver flow capped at 7 Gbps and the New York flow
    capped to 3 Gbps.
    Unfortunately, we still had trouble with the eastern bottleneck, limiting
    New York's rate to 2.5 Gbps.}
    \label{tab:exp_results_diff}
\end{table}

\subsection{Summary}

The results demonstrate four things.
First, for a single flow we can achieve near bottleneck rates while also
maintaining a smooth RTT.
Second, when we specify rate caps we can get smooth RTTs that meet these rates.
Third, when multiple flows with the same RTTs are present, and the flows are
uncapped, the flows  approximately share the bandwidth.
Fourth, when multiple flows with different RTTs are present, and the flows are
uncapped, one flow will dominate.

%% file: buffer.tex
\section{Impact of Buffer Size}

This section was adapted from a paper presented at the 2019 Buffer Conference.
The paper was never officially published, so we present the results here.

\subsection{Simulation Methodology}

To study the impact of buffer size on the performance of the algorithm we used the same discrete-event simulation model as before.
The network is modeled with two buffers; one to model the bottleneck link, and the other for the returning ACKs.
The simulator allows separate client processes representing end systems to be initiated  each running their own instance of the MPC algorithm, which is near identical to the Linux kernel implementation.
Three types of events  namely,  enqueue, dequeue, and ACK are modeled.
Enqueue represents a client sending a packet and enqueueing it on the bottleneck link's buffer.
The enqueue events are specific to each client; the event being scheduled at time intervals determined by the rate set by the MPC algorithm and the size of each packet.
If an enqueue occurs and the bottleneck link buffer is full, a loss is recorded.
Dequeue corresponds to the bottleneck link processing and transmitting the next packet in its queue.
The dequeue events are scheduled at time intervals determined by the rate set by the bottleneck rate and the size of each packet.
ACK is scheduled to occur after one RTT propagation delay from dequeue time, with a small  added noise sampled from a bounded (negative) exponential distribution with mean 1\% and a max of 10\% of the base RTT.
When an ACK is received the RTT is calculated and passed to the MPC algorithm, which updates it's sending rate.

\subsection{Results}

To evaluate the impact of buffer size  we simulated different
numbers of flows that shared a 40~Gbps bottleneck link.
We changed the bottleneck link's buffer size to varying percentages of the
bandwidth-delay product (BDP).
Of the 100,000,000 data points we sub-sampled 1,000,000.
Unless stated otherwise, all plots correspond to a system with two simultaneous
flows running.
For the box plots in this paper, we considered outliers to be points 1.5xIQR beyond the first or third quartile.

\begin{figure}[t]
    \centering
    \includegraphics[width=\linewidth]{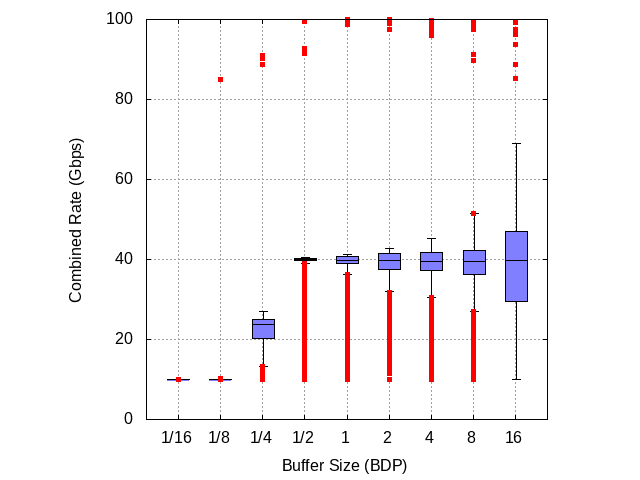}
    \caption{Combined rates for 2 flows versus buffer size in terms of BDP.
    Rates are most stable when the buffer size is at the BDP.
    Decreasing the buffer size far below the BDP limits the rate below the bottleneck rate.
    The high number of outliers near the bottom and top correspond to probing
    for $l_P$ and $l_B$, respectively.
    This accounted for less than 0.01\% of observations, but out of the 1,000,000 observations we get approximately 100 outliers.
    All boxes have a lower limit of 10~Gbps because of each flow was set for a minimum pacing of 5~Gbps (two flows thus always use 10~Gbps).}
    \label{fig:rate}
\end{figure}

\begin{figure}[t]
    \centering
    \includegraphics[width=\linewidth]{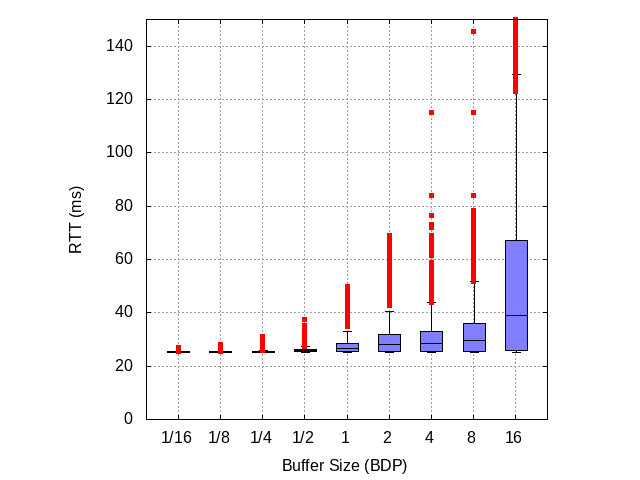}
    \caption{The RTT as a mean among the flows compared to buffer size in terms of BDP.
    The median and range in RTT increase as buffer size increases.
    The large outliers correspond to probes for $l_B$.
    All boxes have a lower limit of 25ms because that is the minimum RTT $l_P$.}
    \label{fig:rtt}
\end{figure}

\begin{figure}[t]
    \centering
    \includegraphics[width=\linewidth, height=2.0in]{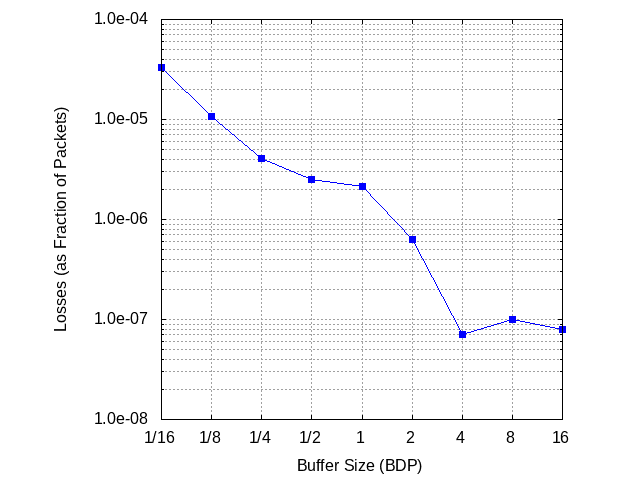}
    \caption{Total losses among all flows  as a function of the  buffer size  in terms of BDP.
    Losses are in logarithmic scale as a fraction of total number of packets.
    Note also that buffer size is logarithmic in base 2, so the curve is not an
    exponential decay.
    We see that losses decrease with increasing buffer size.
    At about 4~BDP losses level out, indicating larger buffers may not improve performance.}
    \label{fig:losses}
\end{figure}

\begin{figure*}[t]
    \centering
    \begin{subfigure}{0.5\textwidth}
        \centering
        \includegraphics[width=\textwidth, height=2.0in]{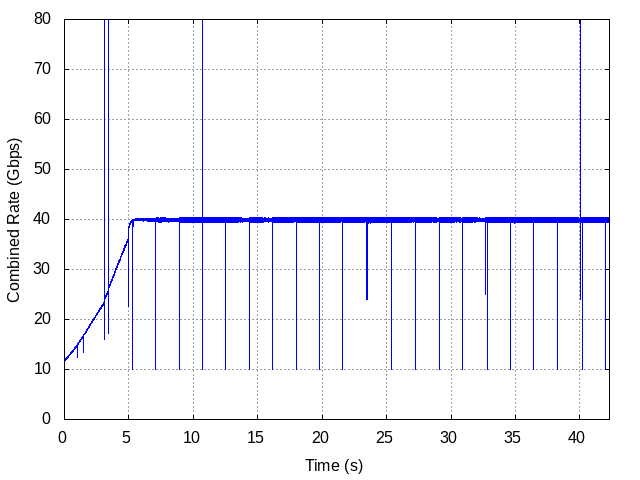}
        \caption{1/2 BDP}
    \end{subfigure}%
    \begin{subfigure}{0.5\textwidth}
        \centering
        \includegraphics[width=\textwidth, height=2.0in]{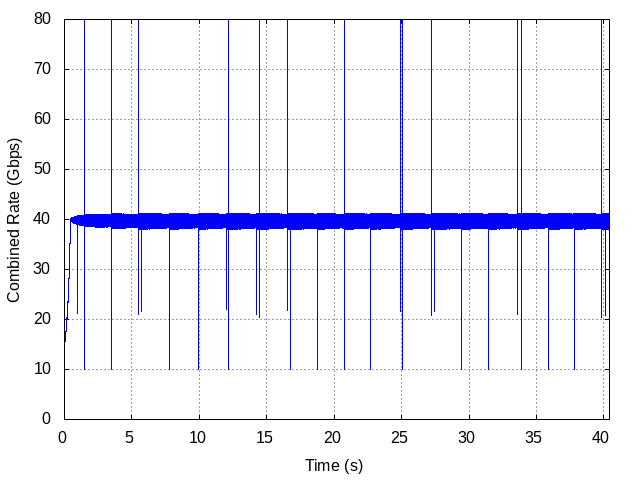}
        \caption{1 BDP}
    \end{subfigure}
    \caption{Plot of combined rate for multiple flows vs time.
    A buffer size of 1/2 the BDP takes about 5s to reach the bottleneck rate, while a size of exactly the BDP takes a fraction of a second.
    Generally, higher buffer sizes allow the algorithm to reach the bottleneck rate faster.}
    \label{fig:op_points}
\end{figure*}

Figure~\ref{fig:rate} shows the effect of changing the buffer size on the achieved rate.
For buffer sizes less than 1/2 the BDP, the throughput rate is limited to less
than the bottleneck rate. This is due to  two reasons.
First, with low buffer sizes there are a large number of losses (see Figure
\ref{fig:losses}), which causes it to back-off more frequently, as the algorithm probes for $l_P$
when a loss occurs in order to prevent future losses.
Second, for small buffer sizes, since the RTT range is more limited (see Figure
\ref{fig:rtt}), the algorithm is limited in the control actions it can take and
is thus more cautious.
As buffer sizes increase beyond 1/2 the BDP, the variation in the rate increases.
This is because as the range in RTT increases, the algorithm must take greater
control action to meet the target RTT (which is halfway between the minimum and
maximum RTT).
The red squares correspond to outliers.
The outliers below the box plot correspond to the algorithm probing for the
minimum RTT, where it decreases the rate to empty the link buffer, which in
turn decreases the RTT.
The outliers above the box plot correspond to the algorithm probing for the
maximum RTT, where it increases the rate to fill the link buffer, which in
turn increases the RTT.
One may notice that when probing for a maximum RTT the pacing rate goes beyond
the bottleneck rate.
This is acceptable because it only occurs for a short time span.
We can see this in Figure \ref{fig:op_points}, where the spikes correspond to
probing.
The probing also immediately stops when a loss is detected.
The figure also demonstrates one advantage the MPC based controller has over traditional AIMD based controllers.
Because the control acts as a continuous function, we can produce a low variance rate for moderate ($\sim$ 1 BDP) buffer sizes that tracks
very close to the bottleneck rate.
This is as opposed to AIMD controllers, whose pacing rate (a function of the window size) exhibits a saw tooth pattern, leading to high variation in the rate.

Figure \ref{fig:rtt} shows the effect of changing the buffer size  on the RTT.
We can see that as buffer size increases so does the range of RTTs.
This corresponds to the core idea of our algorithm that fuller buffers have
longer RTTs.
The explanation for this is that packets must wait for the ones in front of them
to be transmitted first before they can be transmitted, and thus the packets' RTTs increase.
Note that since the RTT set by our control algorithm is a function of the RTT
range, we will find that the median RTT increases with buffer size as well.
The outliers again correspond to probing by the algorithm.
The lower end of the boxes end abruptly at 25 ms because that is the 
propagation delay (i.e., the minimum RTT).
As with the rate, at a reasonable buffer size the algorithm can maintain a low variance RTT
and avoid the saw tooth pattern of AIMD protocols.

Figure~\ref{fig:losses} shows the effect of changing the buffer size on the number of losses.
As expected, increasing the buffer size decreases the number of losses.
What is important to note is that as the buffer size approaches 1 BDP the losses
start to level out, but once it passes 1 BDP the losses drop off rapidly.
This indicates that decreasing the buffer size slightly below 1 BDP, such as to
1/2 BDP, may be worth the trade-off for losses.
At around 4 BDP the losses level off, which is due to the extremely low number
of losses (7-10 out of 100,000,000 packets).

\begin{figure}[t]
    \centering
    \includegraphics[width=\linewidth, height=2.0in]{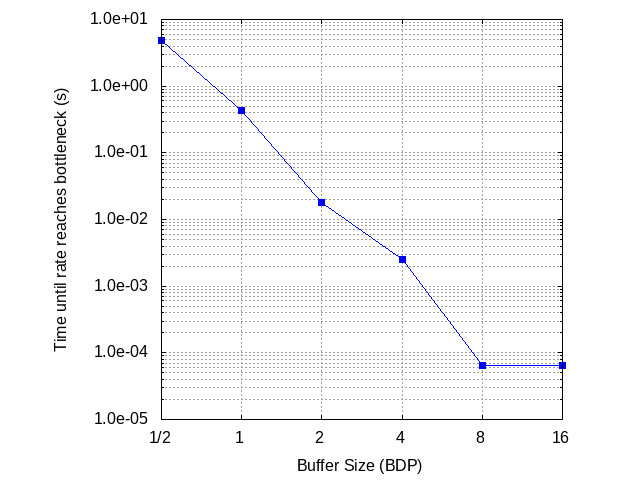}
    \caption{As BDP increases, the time it takes for the algorithm to reach the bottleneck rate decreases.
    This decrease if very rapid, as indicated by the logarithmic scale used.
    At 8 BDP the time to reach the bottleneck rate plateaus, as it reaches the minimum update time for the algorithm (i.e. one RTT).}
    \label{fig:stability}
\end{figure}

A summary of how the rate, RTT, and losses change with respect to the number of
flow and buffer size is shown in Table \ref{tab:summary}.
The pacing rate is largely unaffected by the change in number of flows, and is
mainly a function of buffer size.
The RTT, in contrast, varies with both.
An explanation for this is that when the number of flows increase each individual
flow only accounts for the RTT of its packets, thus they each observe a higher
RTT because they must wait for the other flows' packets.
Similarly, each flow sends at a rate of
$(\text{combined rate})/(\text{\# of flows})$
and thus views a lower rate as the number of flows increases, but when viewing
them together the combined rate does not change.
In short, RTT changes with the number flows because it is a per flow measurement,
while the combined rate does not because it is not measured per flow.
Losses predictably increase with the number of flows and decrease with larger
buffer sizes.
\begin{table*}[t]
    \centering
    
    \begin{tabular}{|c|c|c|c|c|c|c|c|c|c|}
        \hline
        \multirow{2}{*}{Flows} & \multicolumn{9}{|c|}{Buffer Size (BDP)}\\
        \cline{2-10}
        & 1/16 & 1/8 & 1/4 & 1/2 & 1 & 2 & 4 & 8 & 16\\
        \hline
        \multicolumn{10}{|c|}{Median Combined Rate (Gbps)}\\
        \hline
        1 & 10.0 & 10.0 & 33.7 & 39.9 & 39.8 & 39.7 & 39.6 & 39.4 & 39.4\\
        2 & 10.0 & 10.0 & 23.7 & 39.9 & 39.8 & 39.7 & 39.6 & 39.5 & 39.7\\
        4 & 10.0 & 10.0 & 23.5 & 40.0 & 40.0 & 39.8 & 39.8 & 39.7 & 40.3\\
        8 & 10.0 & 10.0 & 24.0 & 39.2 & 40.0 & 40.0 & 40.0 & 40.0 & 39.9\\
        \hline
        \multicolumn{10}{|c|}{Median RTT (ms)}\\
        \hline
        1 & 25.2 & 25.2 & 25.3 & 25.8 & 26.6 & 26.9 & 27.4 & 28.6 & 37.9\\
        2 & 25.2 & 25.2 & 25.3 & 25.7 & 26.7 & 28.0 & 28.4 & 29.7 & 39.2\\
        4 & 25.3 & 25.3 & 25.3 & 25.7 & 26.7 & 28.7 & 29.4 & 30.6 & 42.6\\
        8 & 25.3 & 25.3 & 25.4 & 25.8 & 26.9 & 29.2 & 30.4 & 32.0 & 53.5\\ 
        \hline
        \multicolumn{10}{|c|}{Losses (as Fraction of Packets Sent)}\\
        \hline
        1 & 3.2e-5 & 3.1e-6 & 9.9e-7 & 6.4e-7 & 2.8e-7 & 2.0e-8 & 1.0e-8 & 2.0e-8 & 2.0e-8\\
        2 & 3.3e-5 & 1.1e-5 & 4.1e-6 & 2.5e-6 & 2.1e-6 & 6.3e-7 & 7.0e-8 & 1.0e-7 & 8.0e-8\\
        4 & 4.2e-5 & 2.7e-5 & 1.1e-5 & 7.0e-6 & 5.9e-6 & 3.9e-6 & 2.0e-6 & 2.5e-7 & 2.1e-7\\
        8 & 1.0e-5 & 7.7e-5 & 2.7e-5 & 1.7e-5 & 1.6e-5 & 1.2e-5 & 7.4e-6 & 5.0e-6 & 7.7e-7\\
        \hline
    \end{tabular}

    \caption{The pacing rate and RTT are mainly affected by buffer size.
    Changing the number of flows does not have a significant effect on pacing rate.
    Note that the rate of 40.3 Gbps, which is over the 40 Gbps bottleneck rate, is a result of the high buffer size and many competing flows.
    On the other hand, losses are a function of both the number of competing flows and the buffer size.
    Predictably, an increase in number of flows leads to an increase in losses,
    while increasing buffer sizes decrease the number of losses.}
    \label{tab:summary}
\end{table*}

Results seem to indicate that a buffer size 1/2 the BDP is ideal.
This buffer size leads to a minimal increase in RTT, achieves maximum
throughput, has a low median RTT, and has the lowest variance in both throughput
and RTT.
However, one thing these graphs do not show is how long the flows take to reach the bottleneck rate.
If we look at Figure \ref{fig:stability} and \ref{fig:op_points} we can see how
changing the buffer size affects how long it takes to reach the bottleneck rate.
In general increasing the buffer size causes the algorithm to reach the bottleneck rate faster.
This is because, as mentioned earlier, it does not need to be as cautious to avoid losses.
A size of 1 BDP may be preferable because it reaches the bottleneck rate faster.
In general smaller buffers (1/2 BDP) are preferable for long lived flows, where the long startup time is offset by the increased stability.
On the other hand, larger buffers allow the algorithm to quickly react at the expense of stability, which is preferable for short lived flows.

%% file: relatedwork.tex
\section{Related Work}
\label{sec:relatedwork}

Model predictive control is used in many engineering fields. According to
\citet{bordons2007camacho} MPC is usually used for engineering factories and
processing pipelines. While the book details many 
different approaches for implementing MPC, we primarily  borrow concepts presented in the sections on
``State Space Formulation'' and ``Predictive Functional Control''
\cite{bordons2007camacho}. 

In the development of our MPC based control algorithm, we borrow a few concepts from BBR TCP \cite{bbr_tcp}. In particular we use the concepts of a network's `propagation
delay' and `bottleneck rate.' However, beyond these concepts and the use of RTT
for feedback, BBR and our work are very different in terms of their control
mechanisms. Other notable TCP congestion control algorithms are CUBIC \cite{cubic} and HTCP
\cite{htcp}. In general CUBIC aims to be a general purpose congestion control
that quickly adjusts its window without over-saturating the line. HTCP aims to
be a fair congestion control algorithm that tries to rapidly use as much
bandwidth as possible without crowding out other flows, and in general be useful
for high-speed, long-distance networks.

The Fair Queuing (FQ) and FQ CoDel \cite{linux_kernel} qDiscs were used heavily
as a reference point in our research. FQ tries to balance packet flow loads by
using per flow pacing, and by dequeuing packets in round robin order. FQ CoDel extends FQ with the CoDel AQM scheme to provide stochastic
classification and to avoid bufferbloat. One notable advantage of FQ over FQ
Codel is that provides flow pacing to applications and higher-level congestion
control algorithms. This is why FQ is suggested to be used with BBR and not FQ
Codel.

There is limited work in applying MPC to
congestion control. The study in   \cite{humaid} used a window based MPC
control that leveraged random early detection (RED). They employ a non-linear
model that is optimized via selection from a three dimensional array of PID
controllers \cite[46,48]{humaid}. Their simulated results show that they can
closely match queue lengths to a desired value.

Our work extends \citet{david} which also developed an MPC based control algorithm that minimized RTT and its variance, while at the same time
maximized the pacing rate. A key limitation was that  the model was not tied directly to the network,
which made it brittle and difficult to adapt. Furthermore, minimizing the 
RTT while maximizing pacing rate led to a large amount of volatility as they
were in conflict. We have attempted to fix these issues in our current design.

%% file: conclusion.tex
\section{Conclusion}
\label{sec:conclusion}

In this paper we present the design, implementation, and evaluation of a  congestion control algorithm based on MPC.  Employing MPC on a pacing rate control and an RTT response allows for a per acknowledgment, real time feedback system.
This enables us to achieve rate and RTT stability not seen in many other TCP
congestion control algorithms.
It has the potential to allow for more stable control actions than loss based
methods.
This is because losses happen in bursts, and do not allow for
real-time response like RTT based methods.

The Linux kernel modules were developed for the congestion control and qDisc
layers.
This allows users to test and use our algorithm on any existing Linux system.
Allowing for rate setting per flow increases the usefulness of these modules.
Users can use this to give priority to certain flows.
In science networks such as ESnet efforts are underway to develop a
high-precision network telemetry tool that can provide real-time and
fine-grained network information.
A potential approach could be to enhance the proposed congestion control and
qDisc algorithms using information from such tools.

There are several next steps.
One improvement would be to add an auto-tuning phase to the
implementation, which would save the user time and effort.
As a long term goal we may extend the model to optimize more
than one time step into the future.
This poses challenges to time complexity, but could possibly make the algorithm
more stable.